\documentclass[12pt]{article}
\usepackage{graphicx}
\usepackage{graphics}
\input epsf

\def\beqn{\begin{eqnarray}}
\def\eeqn{\end{eqnarray}}
\relax
\def\ba{\begin{array}}
\def\ea{\end{array}}

\def\beq{\begin{equation}}
\def\eeq{\end{equation}}
\def\bea{\begin{array}}
\def\eea{\end{array}}

\def\[{\left[}
\def\]{\right]}
\def\({\left(}
\def\){\right)}

\def\sm0{{\widetilde{m}_0}}

\def\U1em{{U(1)_{\rm em}}}

\def\sq2{\sqrt{2}}

\def \met {{\,/\!\!\!\!E_{T}}}
\def\End{\end{document}}

\def\sm{{\tilde{m}}} 
\begin{document}

\title{\Large \bf Results on QCD Physics from the CDF-II Experiment}
\date{ }

\author{\large C. Pagliarone\thanks{Corresponding author. E-mail: pagliarone@fnal.gov} \bigskip \\
{\it Universit\'a di Cassino \& INFN Pisa, Italy}\bigskip \\
{(on the behalf of the CDF-II Collaboration)}}
\maketitle

{\large

\begin{abstract}
In this paper we review a selection of recent results obtained, in the area of 
QCD physics, from the CDF-II experiment that studies $p\bar{p}$ collisions 
at $\sqrt{s}$=1.96~TeV provided by the Fermilab Tevatron Collider.
All results shown correspond to analysis performed using the Tevatron Run II data samples.
In particular we will illustrate the progress achieved and the status
of our studies on the following QCD processes:
jet inclusive production, using different jet clustering algorithm, 
$W(\rightarrow e \nu_e)+\,$jets and $Z(\rightarrow e^+ e^-)+\,$jets production,
$\gamma+b-$jet production, dijet production in double pomeron 
exchange and finally exclusive $e^+e^-$ and $\gamma \gamma$ production. 
No deviations from the Standard Model have been observed so far.
\end{abstract}

\vspace{1.pc}
\section{Introduction}   

The Quantum Chromo Dynamics (QCD) processes provide signals to test theoretical calculations
and models and contribute major backgrounds to many other searches or measurements. 
Thus, their detailed understanding and modelling is of crucial importance.
In particular, the current QCD physics program, at Tevatron, includes studies of jets 
with the goal of performing precision measurements to test 
and further constrain the validity of the Standard Model (SM).
Jets can be defined as collimated sprays of particles originating, all in one point, from 
the fragmentation of a parton. The ability in reconstructing the jets allows to characterize and 
measure the energy of the parent partons. As jet calculations, at leading order and at 
higher orders, can vary the definition of a jet it is therefore important in order to compute the jet 
energy beyond the leading order.
Jet energies are measured experimentally by adding the energy of the calorimeter cells associated 
to a cluster, using predetermined algorithms.
During Run~II, CDF-II have been studying alternative methods to the 
fixed cone-based jet clustering algorithm, the so called JetClu~\cite{JetClu} used during 
the Tevatron Run~I (1992-1995). This is needed in order to avoid problems of infrared and 
collinear divergences due to soft partons and below/above threshold particle emission.
Both MidPoint\cite{midpoint} and $k_T$\cite{kt} jet reconstruction algorithms have been used 
in Run~II (see Fig.~\ref{jetdef}). The former is an improved version of the seed cone-based 
JetClu algorithm which reduces the sensitivity to infrared and collinear problems.
The latter starts by finding pairs of nearby particles in the defined phase-space 
and then merges them together to form new pseudo-particles,
continuing until a set of stable well-separated jets are found. This algorithm is infrared and collinear safe 
to all orders in perturbative QCD (pQCD).
In addition to the energy from the primary parton, jets accrue soft contributions from the underlying 
event (UE) of beam remnants. 
These contributions become more important at smaller jet $p_T$~\cite{cdf_cord}.
During Run~I and Run~II, the contribution from UE energy has been studied and a modification 
to Pythia~\cite{pythia} Monte Carlo (MC) has been determined (Tune A~\cite{tuneA}) using CDF-II data. 
Pythia MC with the new set of parameters describes well the jet shapes measured in Run~II~\cite{jetshapes}.
\begin{figure}[t!]
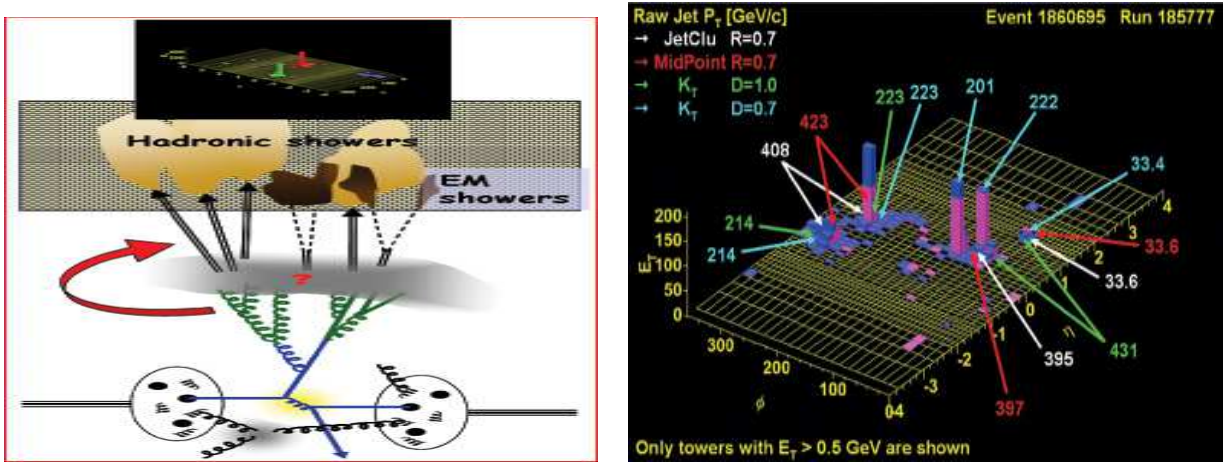

\begin{center}
\includegraphics[width=7.9cm,height=5.9cm]{pict1.eps}\hspace{0.6pc}
\includegraphics[width=7.9cm,height=6.1cm]{pict2.eps}
\caption{ \footnotesize a) Well-defined algorithms for clustering calorimeter towers into 
"jets" is needed in order to make measurements and for comparison with theoretical predictions;
b) Different Algorithms give different results.}
\label{jetdef}
\end{center}
\vspace{-1.7pc}
\end{figure}

\section{The CDF-II Detector}

CDF-II is a $5000$ ton multi-purpose particle physics
experiment~\cite{CDF-1} dedicated to the study of
proton-antiproton collisions at the Fermilab Tevatron collider. It
was designed, built and operated by a team of physicists,
technicians and engineers that by now spans over $44$ institutions
and includes, approximately, more than $500$  members. The history
of the experiment goes back over $20$ years. The CDF detector has
been upgraded~\cite{CDF-upgrades} in order to be able to
operate at the high radiation and high crossing rate of the Run II
Tevatron environment. In addition, there have been several
changes to improve the sensitivity of the detector to specific physics
channels such as heavy flavor physics, Higgs boson searches and many
others. Fig.~\ref{CDF}.a shows an isometric cutaway view of the final
configuration of the CDF-II detector. The central tracking volume
of the CDF experiment has been replaced entirely with new
detectors (see Fig.~\ref{CDF}), the central calorimeters has not been
changed. These upgrades can be summarized as follows:
a new Silicon System  done of $3$ different
tracking detector subsystems: Layer00 installed
directly on the beam pipe, a new Silicon Vertex Detector (SVX II),
an Intermediate Silicon Layer detector (ISL); 
a new central tracker the Central Outer Tracker (COT) that is an open cell drift chamber
able to operate at a beam crossing time of 132~$ns$ with a maximum
drift time of $\sim\!100$~$ns$;
a scintillator based Time-of-Flight detector (TOF), 
new  Plug Calorimeters and an extended upgraded Muon system that 
almost doubled the coverage in the central;
A new Data Acquisition System (DAQ) has also been constructed
in order to operate in the shorter bunch spacing conditions.

\begin{figure}[t!]
\begin{center}
\includegraphics[width=7.9cm,height=6.8cm]{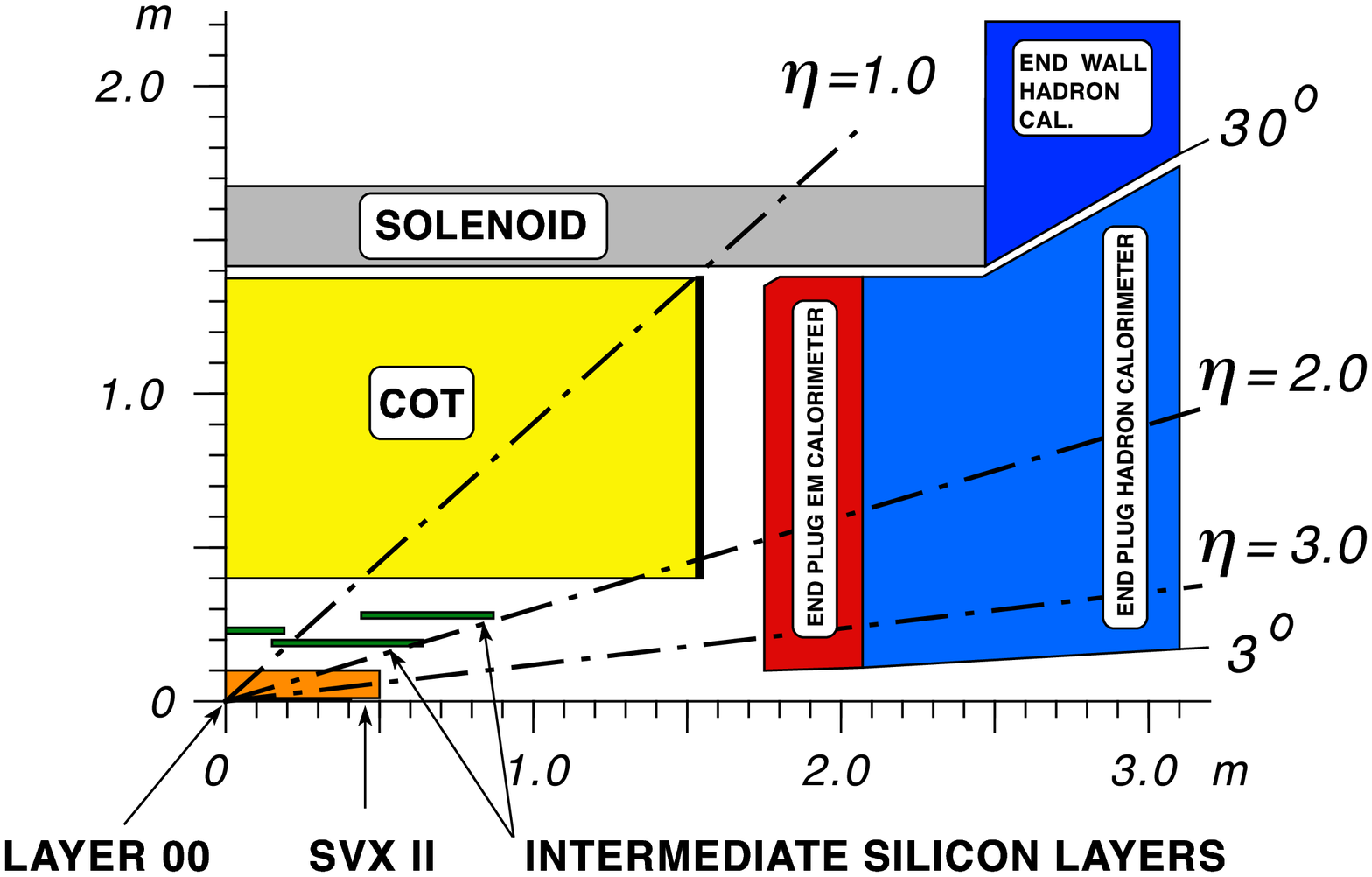}
\hspace{0.6pc}
\includegraphics[width=7.9cm,height=6.8cm]{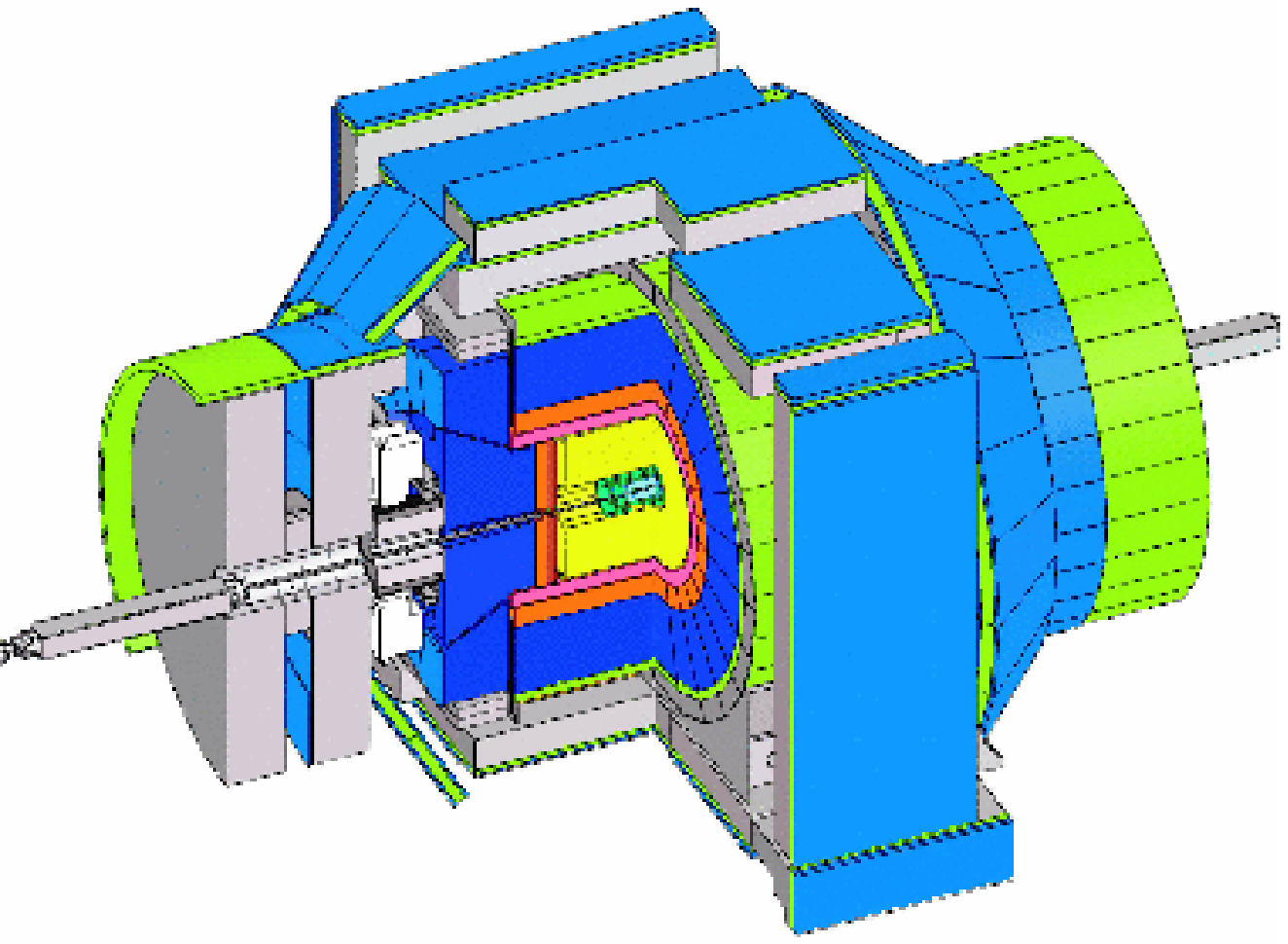}
\vspace{-0.4pc} 
\caption{\footnotesize a) An overview of the Collider
Detector at Fermilab (CDF) in its Run II configuration (CDF-II). b) 
A cutaway view of one quadrant of the
inner portion of the CDF-II detector showing the tracking region
surrounded by the solenoid and end-cap calorimeters.}
\label{CDF}
\end{center}
\vspace{-1.7pc}
\end{figure}

\section{QCD Results from CDF-II}


\begin{figure}[t!]
\begin{center}
\includegraphics[width=7.9cm,height=6.9cm]{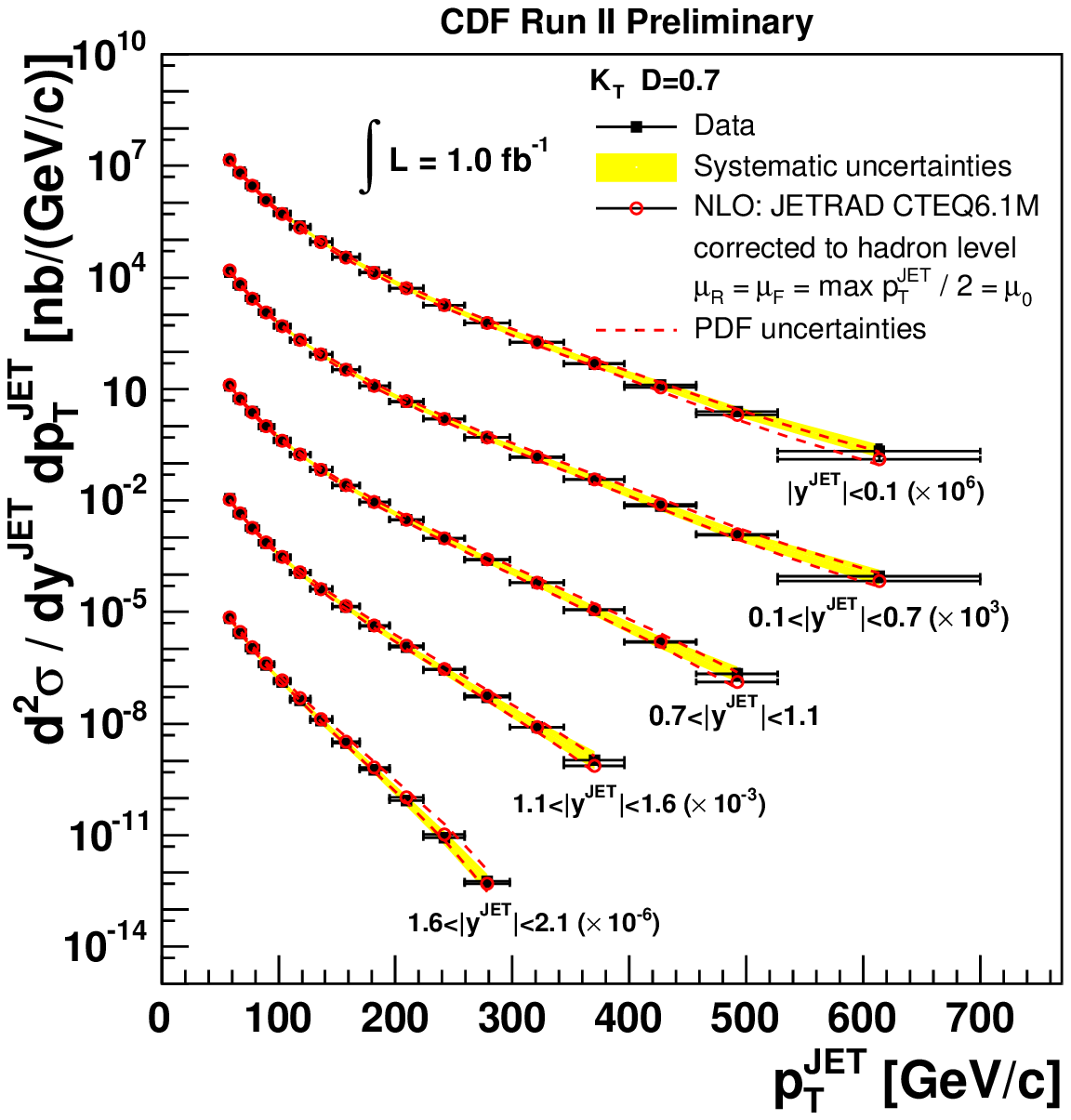}\hspace{0.6pc}
\includegraphics[width=7.9cm,height=6.9cm]{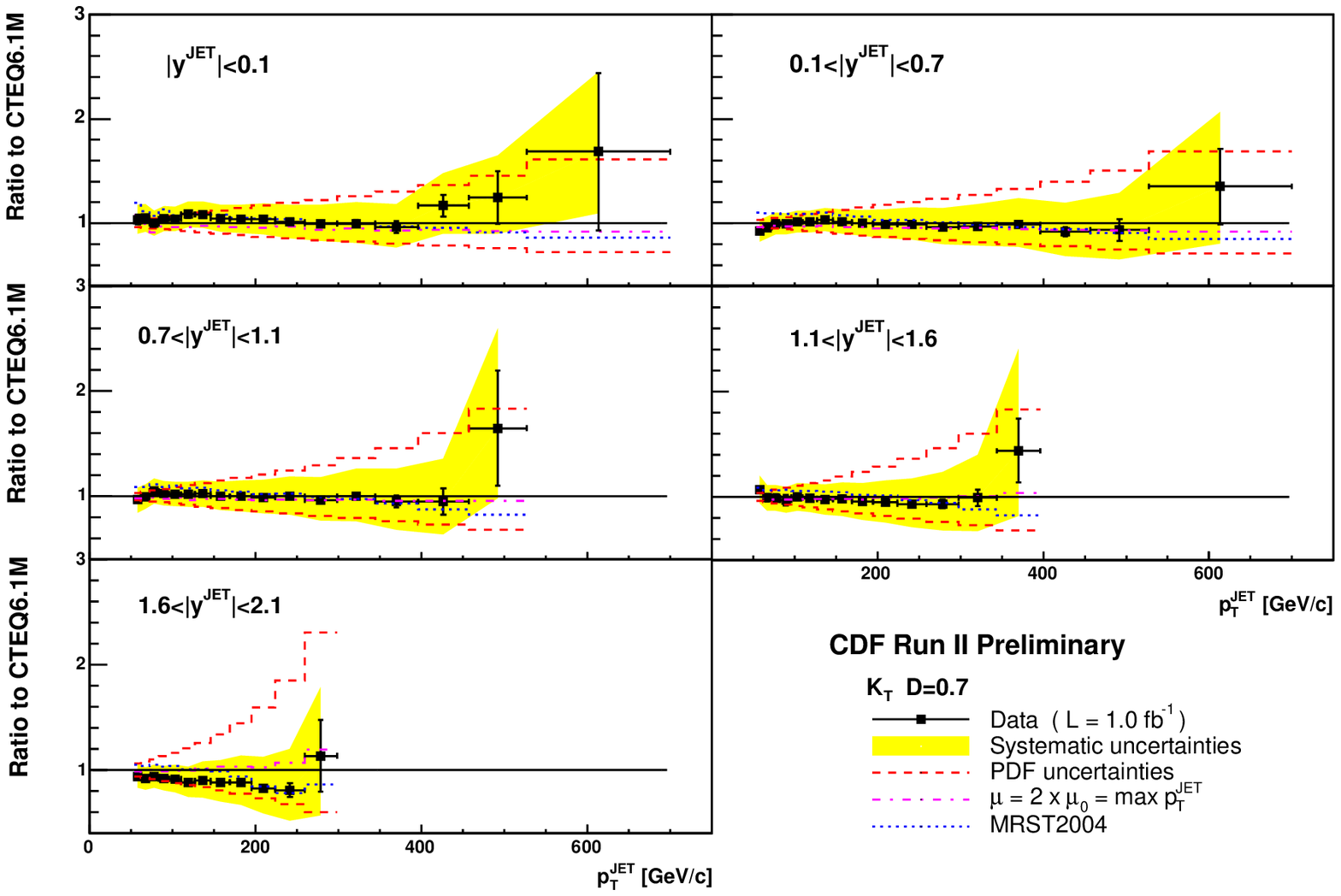}
\vspace{-0.4pc}
\caption{\footnotesize Jet Inclusive Cross section for jets defined using the $K_T$ algorithm.
a) Cross section in five rapidity bins (black dots) as a function of $p^{jet}_{T}$ compared to NLO pQCD predictions (histogram). 
The shaded bands show the total systematic uncertainty on the measurement.
b) Ratio of data to theory as a function of $p^{jet}_{T}$. 
The error bars (shaded band) show the total statistical (systematic) uncertainty on the data.}
\label{JICS1}
\end{center}
\vspace{-1.7pc}
\end{figure}

\subsection{Inclusive jet production: Midpoint and $\mathbf{K_T}$ Analysis}
There are many important reason to study the inclusive jet production at Tevatron.
As matter of fact it is a stringent QCD test up to $8 \div 9$ order of magnitude; 
it is a measurement sensitive to the structure of the parton distribution functions (PDF); 
it helps to add constrains on gluon structure of the PDFs at high$-x$; 
it's a test sensitive to distances up to $10^{-19}$ m and it is an important tool 
for searching for New Phenomena.
The increase in center of mass energy, from 1.80 (Run~I) to 1.96~TeV (Run~II), 
results in a larger kinematical range for measuring the jet production.
The inclusive jet production cross section have been measured at CDF-II 
using two different jet definitions: the $k_T$ algorithm and the MidPoint cone algorithm.
The jets were selected with $p^{jet}_T\ge 54$~GeV/c in five different jet rapidity regions: 
$|y|<0.1$, $0.1<|y|<0.7$, $0.7<|y|<1.1$, $1.1<|y|<1.6$, and $1.6<|y|<2.1$~\cite{cdf_cord}. 
The results shown in this paper are based on $1.04$ fb$^{-1}$ and are recently updated results.
The experimental data are in agreement with next-to-leading order (NLO) calculations (see Fig.~\ref{JICS1} and Fig.~\ref{JICS2}). 
In particular the CDF-II measurements of the inclusive jet cross section, using the $k_T$ algorithm, show that this algorithm works
well in hadron collider environment in the $p^{jet}_{T}$ range studied. 


\begin{figure}[t!]
\begin{center}
\includegraphics[width=7.9cm,height=6.9cm]{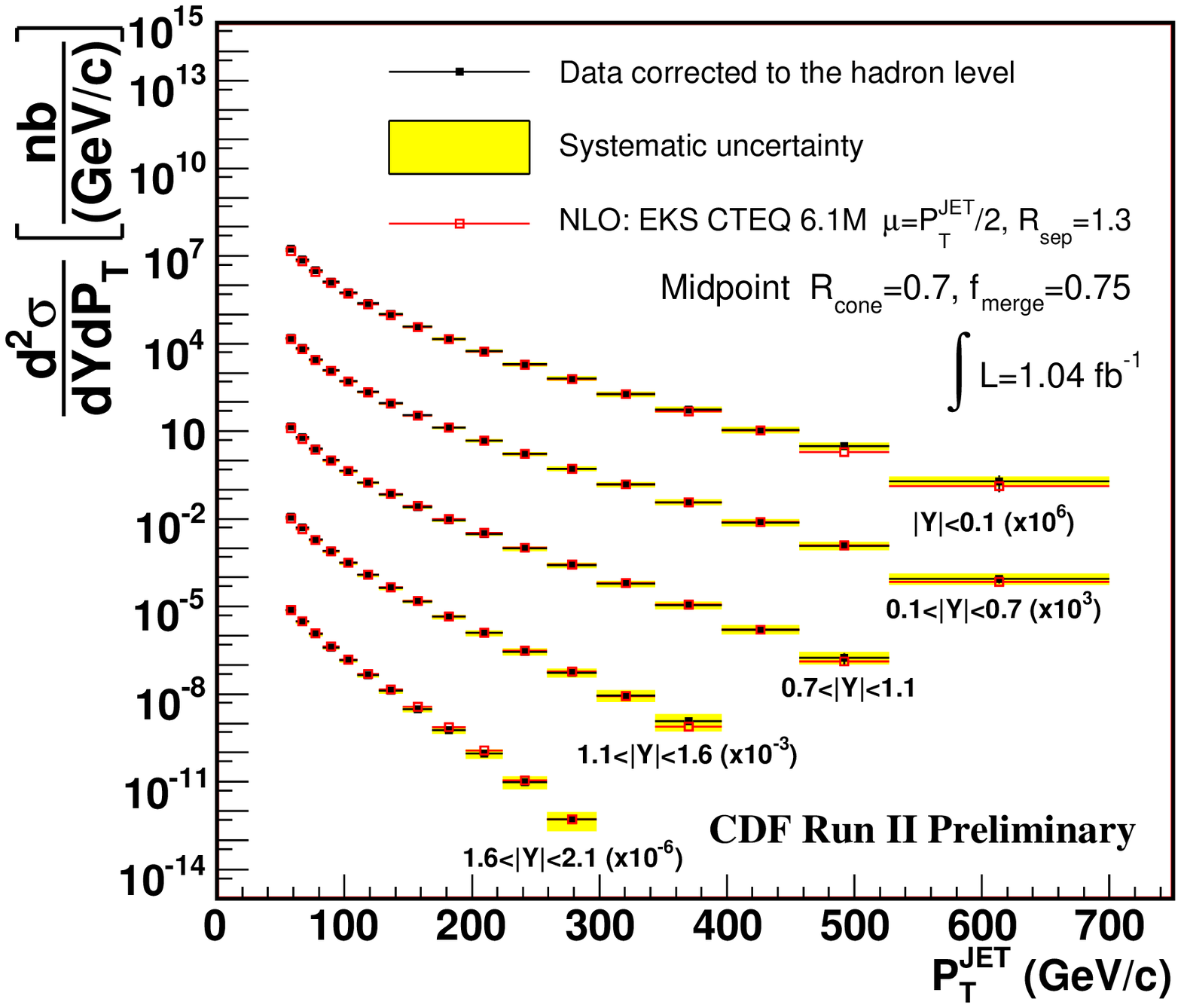}\hspace{0.6pc}
\includegraphics[width=7.9cm,height=6.9cm]{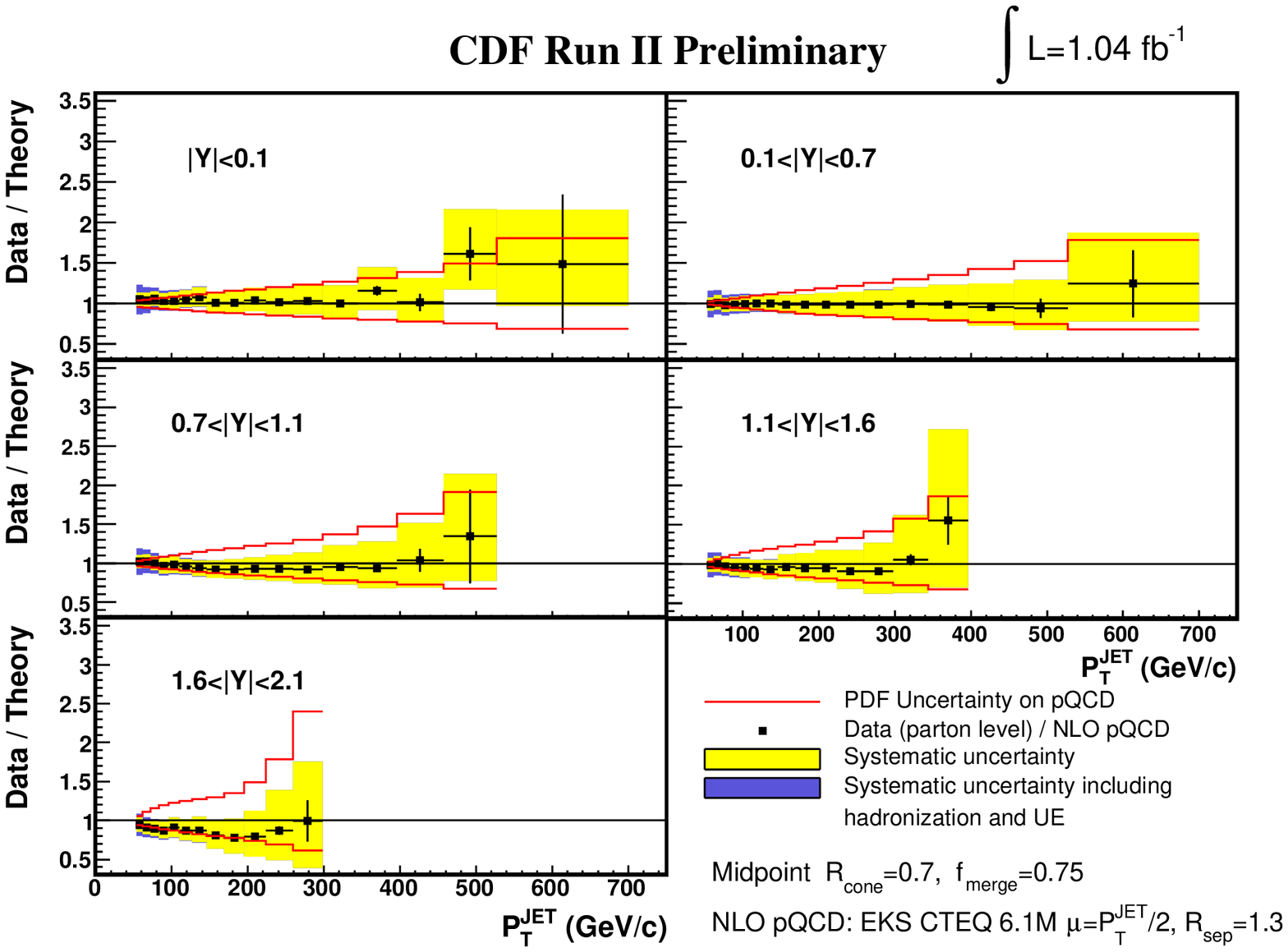}
\vspace{-0.4pc} 
\caption{\footnotesize Jet Inclusive Cross section for jets defined using the MidPoint algorithm.
a) Cross section in five rapidity bins (black dots) as a function of $p^{jet}_{T}$ compared to NLO pQCD predictions (histogram). 
The shaded bands show the total systematic uncertainty on the measurement.
b) Ratio of data to theory as a function of $p^{jet}_{T}$. 
The error bars (shaded band) show the total statistical (systematic) uncertainty on the data.}
\label{JICS2}
\end{center}
\vspace{-1.7pc}
\end{figure}

\subsection{$\mathbf{W/Z}$ boson + jets production}
The production of $W/Z+$jets provides a good test of pQCD, in a multi-jet environment, 
since the presence of the $W/Z$ ensures that the event has a high $Q^2$.
More importantly, $W/Z+\,$jets is a possible signature for many new and important 
processes such as the production of top pairs and single top quark production, the Higgs boson, 
and Supersymmetric particles. 
QCD production of $W/Z+$jets is a large background for many of these searches, and therefore, it is important to 
measure its cross section. QCD Matrix Element (ME) calculations are used to describe the hard scattering in $W/Z+\,$jet 
events, and then Parton Showering Monte Carlo (PS) is used in order to simulate the soft radiation and hadronization. 
An overlap in phase space between $W/Z+n\,$­partons and $W/Z+(n + 1)\,$­partons can lead to double counting when combining 
MC samples to obtain $W/Z+n\,$jets.

\subsubsection{$\mathbf{W(\rightarrow e \nu_e)+\,}$jets production}
CDF-II has measured the $W(\rightarrow e \nu_e)+\,$jets cross section for $W$ plus at least 1, 2, 3 and 4 jets, as a function of the jet 
transverse energy ($E^{jet}_T$).
We also measured the $W(\rightarrow e \nu_e)+\,$jets cross section for events with two or more jets, as a function 
of the dijet invariant mass ($M(j_1,j_2)$), and as a function of the distance in the $\eta-\phi$ plane between 
the leading jets ($\,$$\Delta R \equiv \sqrt{(\phi_{j_1}-\phi_{j_2})^2+(\eta_{j_1}-\eta_{j_2})^2}$$\,$).
\noindent
In order to be model independent, the analysis have been performed in a restricted $W$ kinematics phase space.

\noindent
Events were selected requiring the presence of an isolated electron, in the fiducial rapidity region 
$|\eta(e)|<$ $1.1$ with $E^{e}_{T}>$ $20.0$ GeV, the presence of missing
transverse energy ($\met$) with $\met > 30$ GeV~\cite{cdf_cord}, 
and finally the presence of jets having $|\eta(jet)|<$ $2.0$ and $E^{jet}_{T}>$ $15.0$ GeV.
The reconstructed transverse $W$ mass was required to be $M^{W}_{T}>$ $20.0$ GeV. 
Jets were reconstructed using the JetClu algorithm and corrected at hadron level; no underlying event corrections were applied.
The comparison between data and theoretical calculations have been performed using LO Alpgen MC plus Pythia PS. 

\noindent
The Monte Carlo normalized to the data show a good agreement with the theoretical predictions as shown in
Fig.~\ref{W-jet} and in Fig.~\ref{W-jet-kin}.
 

\begin{figure}[t!]
\begin{center}
\includegraphics[width=7.9cm,height=6.9cm]{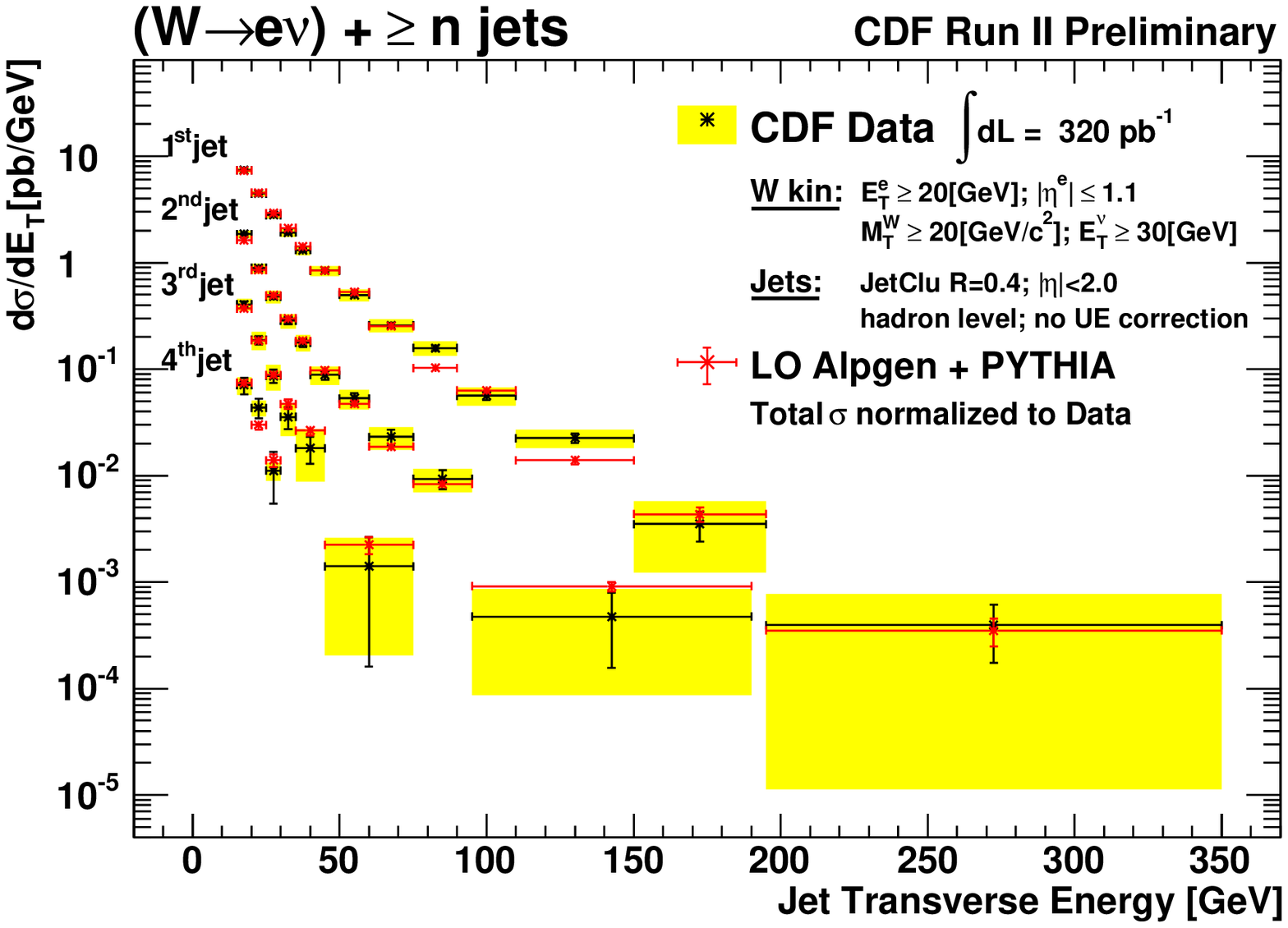}\hspace{0.6pc}
\includegraphics[width=7.9cm,height=6.9cm]{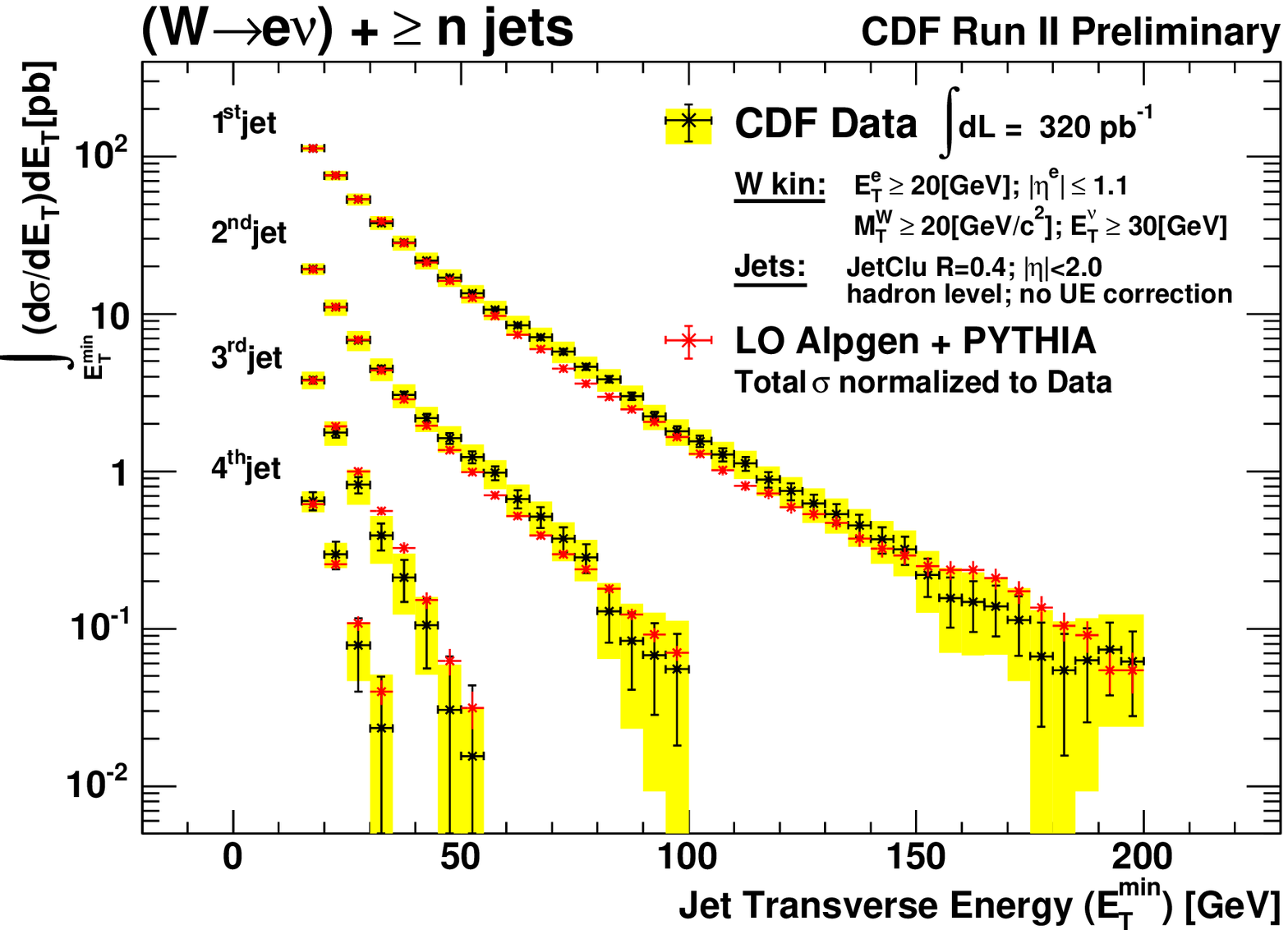}
\vspace{-0.4pc} 
\caption{\footnotesize $W(\rightarrow e \nu_e)+\,$jet production.
a) Differential cross-section for the leading jet in $\geq 1$ jet events, second jet $\geq 2$ jets events, third jet $\geq 3$ and so on.
b) Integrated cross sections from the previous plot. Here the bin is the minimum $E_T$ above which the 
cross section is integrated.}
\label{W-jet}
\end{center}
\vspace{-1.7pc}
\end{figure}

\subsubsection{$\mathbf{Z(\rightarrow e^+ e^-)+\,}$jets production}

Between the QCD analysis, CDF-II is also studying the $Z(\rightarrow  e^+ e^-)+\,$jets production. As matter of fact 
a precise measurement of $Z(\rightarrow  e^+ e^-)+\,$jets cross section is fundamental in 
order to estimate the  $Z(\rightarrow  \nu_e \bar{\nu}_e)+\,$jets irreducible background.
At the present we are looking for a Monte Carlo simulation that properly describes
the final event topology. In Fig.~\ref{Z-jet}.a we show the comparison of the jet production, 
as a function of the $p^{jet}_T $, in the data and in Pythia Tune A Monte Carlo simulation.
Fig.~\ref{Z-jet}.b gives the differential shape of the jets, plotted in steps 
of $\Delta R=$ $0.1$, using the calorimeter towers. 
Jets have been reconstructed using the MidPoint Algorithm with R = 0.7. Events have been selected requiring only one reconstructed 
vertex in the event. This analysis is in progress further results will come soon.

\subsection{$\mathbf{\gamma+b-}$jet production}

CDF-II is searching, starting from Run I, for events containing $\gamma+b$-jet. 
This class of analysis is interesting both for QCD studies both for searching for new phenomena, 
in particular, in the light stop scenario or looking for techniomega production. 
The search, that we present here, is based on $340$ pb$^{-1}$ of data, collected requiring the presence,
in the events, of one isolated $\gamma$, in the fiducial rapidity region $|\eta(\gamma)|<$ $1.1$
with $E^{\gamma}_{T}>$ $26.0$ GeV, one $b$-jet (a positively signed displaced secondary vertex) with $|\eta(b)|<$ $1.5$,
$E^{b}_{T}>$ $20.0$ GeV and $\Delta R(\gamma,b) \equiv  \sqrt{(\phi_{\gamma} - \phi_{b})^2 +
(\eta_{\gamma} - \eta_{b})^2}>$ $0.7$.

\begin{figure}[t!]
\begin{center}
\includegraphics[width=7.9cm,height=6.9cm]{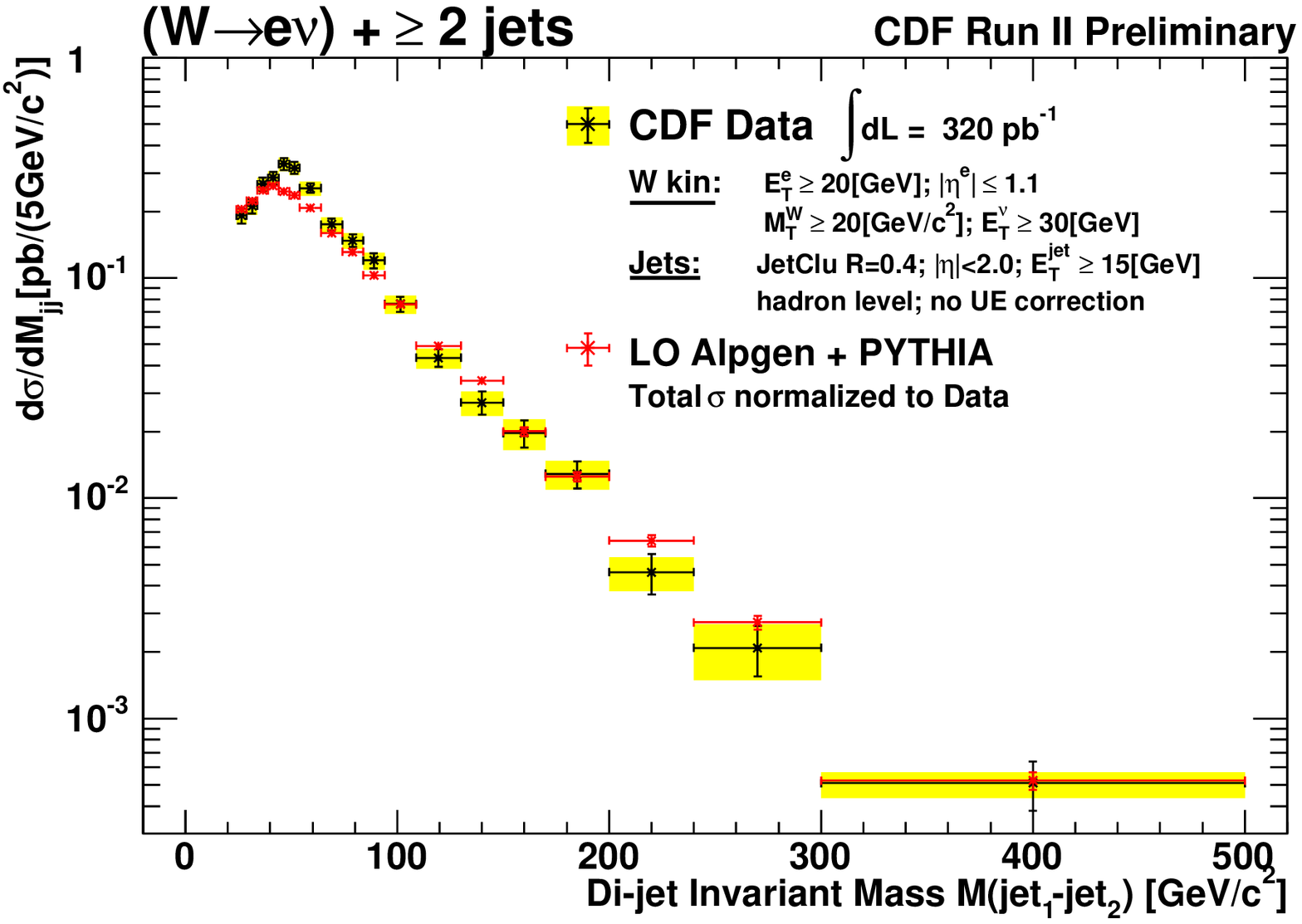}\hspace{0.6pc}
\includegraphics[width=7.9cm,height=6.9cm]{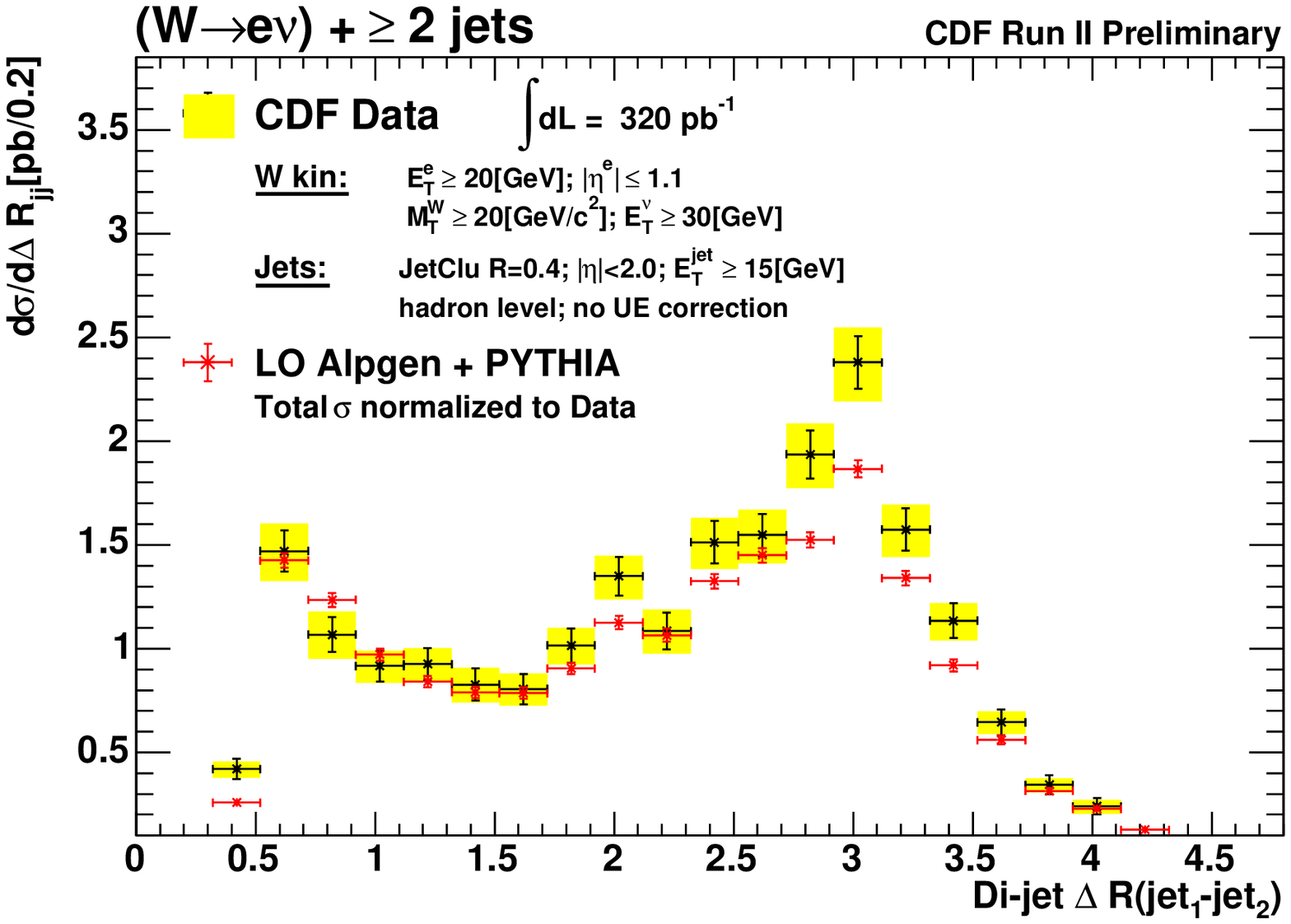}
\vspace{-0.4pc} 
\caption{\footnotesize $W(\rightarrow e \nu_e)+\,$jet kinematics
for $W+ \geq 2\,$jets, where both jets have a minimum $E^{jet}_T>15$ GeV.
 a) First-second jet invariant mass differential cross section.
 b) First-second jet $\Delta R$ differential cross section.}
\label{W-jet-kin}
\end{center}
\vspace{-1.7pc}
\end{figure}

\noindent
We fit the secondary vertex mass in the data in order to determine the $b-$jet fraction. 
To estimate the background from fake $\gamma+b-$jet we use preshower detector information, 
to calculate the number of fake photons in our sample, and  we multiply this by the $b-$jet 
fraction, previously evaluate usisng a representative background data sample. 
The calculated cross-section is given in the table below, where we quote the differential and 
inclusive cross-sections for $\gamma+b-$jet production, within 
the kinematical range specified in the table, given as a function of photon transverse energy ($E_T^\gamma$). 
The first error shown is statistical, the second is the systematic uncertainty. 
\begin{figure}[h!]
\begin{center}
\includegraphics[width=8.4cm,height=4.5cm]{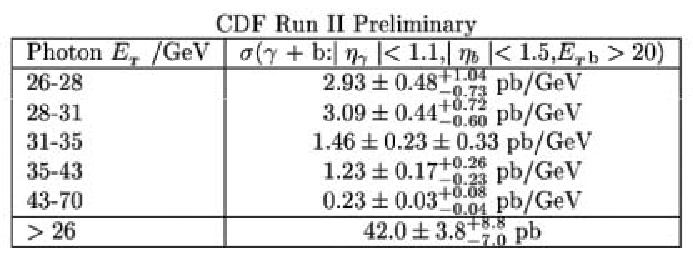}
\end{center}
\vspace{-1.7pc}
\end{figure}


\subsection{Dijet production in DPE}

One of the most important question, in hard diffractive processes, is whether or not 
they obey QCD factorization. In other words, whether the pomeron has a universal, 
process-independent PDF. Results on diffractive Deep Inelastic Scattering (DIS) from the 
$ep$ collider HERA show that QCD factorization holds in DIS.
\begin{figure}[t!]
\begin{center}
\includegraphics[width=7.9cm,height=6.9cm]{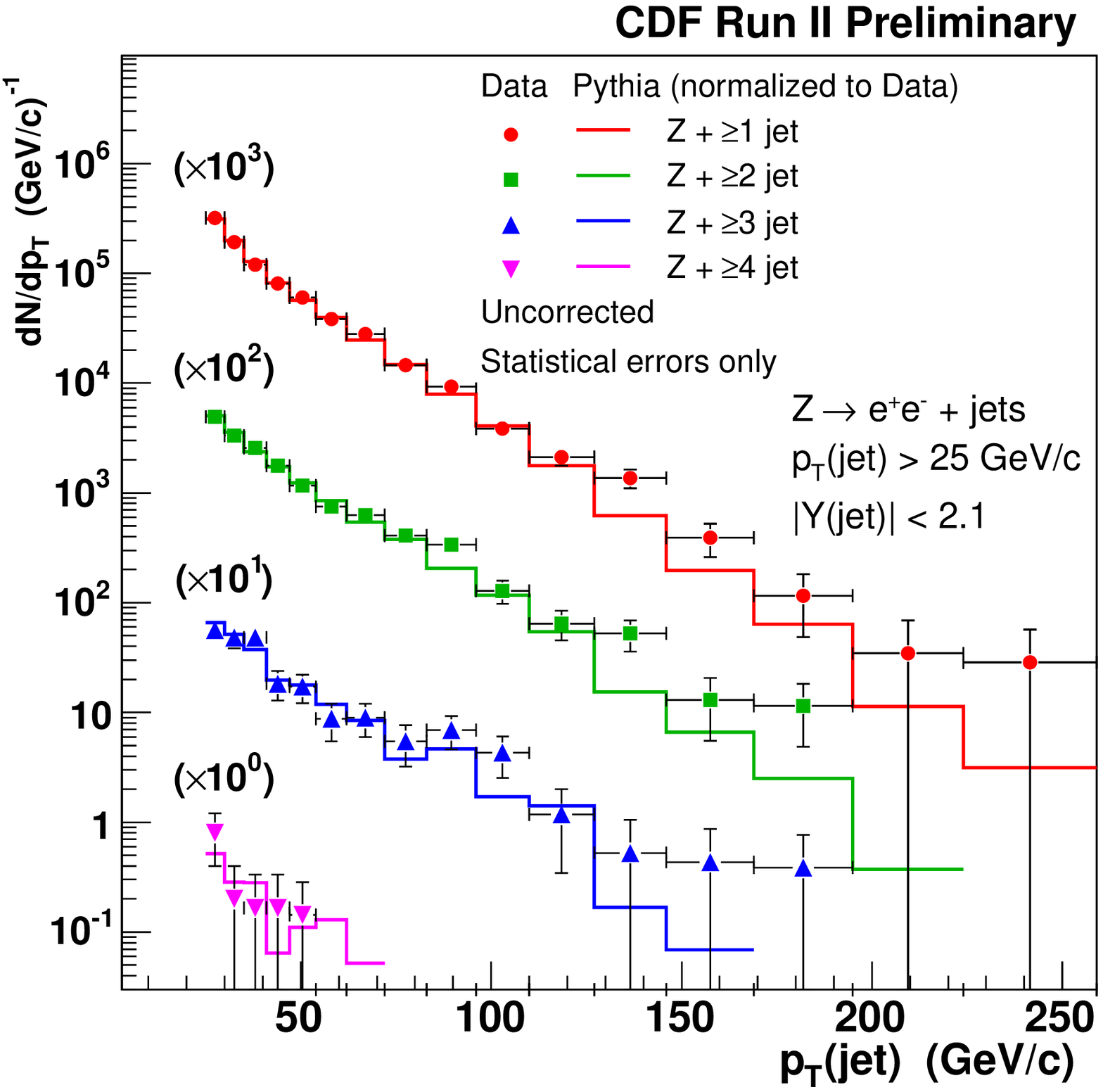}\hspace{0.6pc}
\includegraphics[width=7.9cm,height=6.9cm]{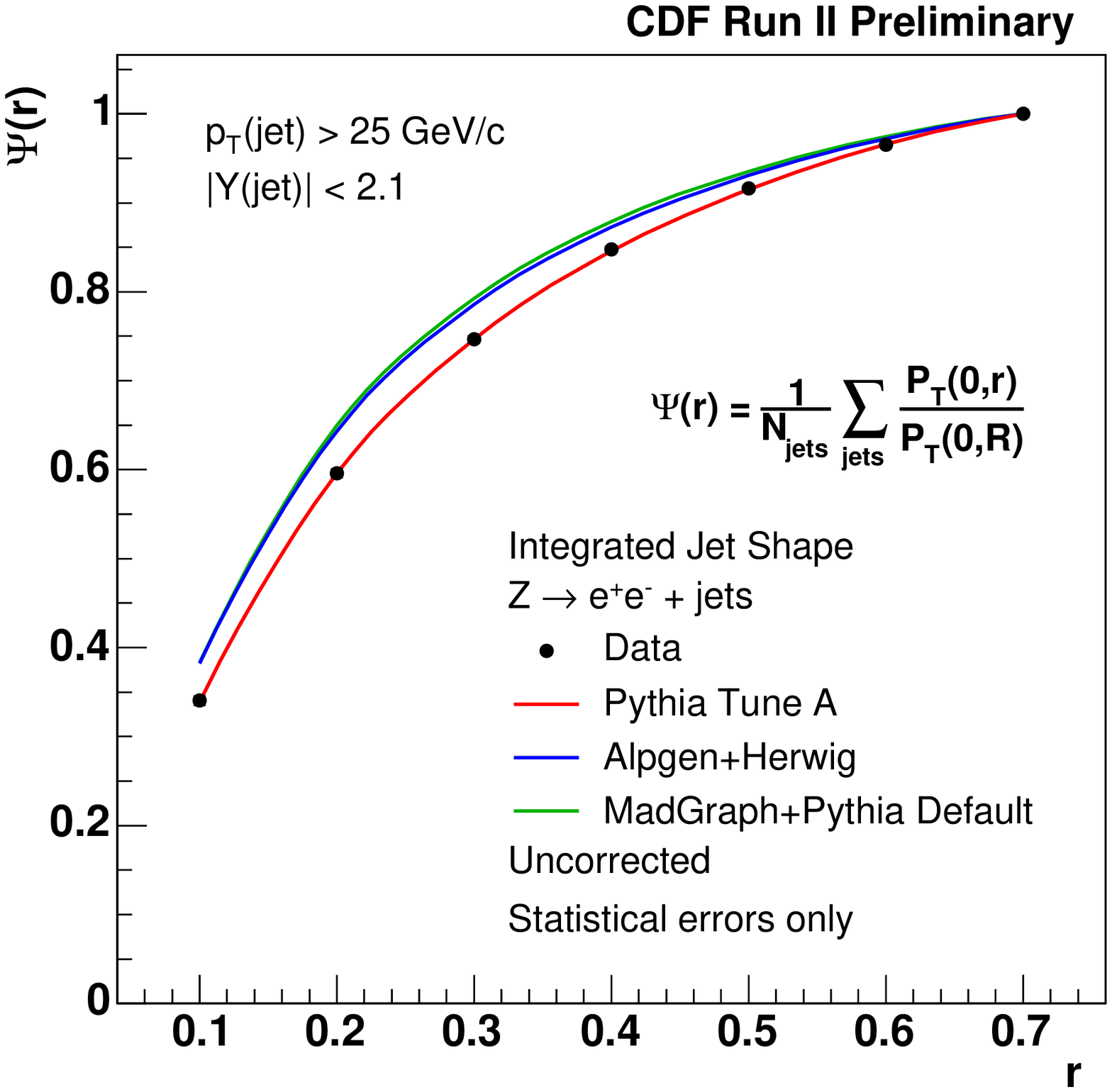}
\vspace{-0.4pc} 
\caption{\footnotesize $Z(\rightarrow  e^+ e^-)+$jets production. 
a) Comparison of the inclusive jet production, as a function of the $p^{jet}_T $, in the data and in Pythia Tune A MC simulation.  
b) Integrated shape of the jets, plotted using the calorimeter towers. Jets are reconstructed with MidPoint algorithm 
with R = 0.7. The event selection requires only one reconstructed vertex in the event.}
\label{Z-jet}
\end{center}
\vspace{-1.7pc}
\end{figure}
Severe breakdown of QCD factorization in hard diffraction between Tevatron and HERA data
have been observed. Single Diffractive (SD) process rates of dijet, $W$-boson, $b$-quark, and  $J/\Psi$  
production relative to non diffractive (ND) ones, measured in Run I at CDF, are about an order of 
magnitude lower than expectations from PDFs determined at HERA.

\noindent
The present CDF-II analysis is based on $310$ pb$^{-1}$ of data, collected with dedicated 
Diffractive Triggers, triggering on a leading antiproton in the Roman Pots (RP) in conjunction 
with a least one jet in the calorimeters. ND dijet events have been selected triggering only on jet requirements.
We observe an excess of data events over the backgrounds obtained from inclusive DPE dijet Monte Carlo simulation.
The observed excess is consistent with exclusive dijet Monte Carlo predictions in terms of kinematical distribution shapes.
The cross sections of the observed exclusive dijet events are measured using a combination of inclusive DPE and 
exclusive dijet Monte Carlo simulations.
The results are summarized in Fig.~\ref{DPE}. In particular Fig.~\ref{DPE}.a show the diffractive dijet exclusive cross 
section compared with the hadron level predictions, coming from the ExHuME MC simulation, and with the exclusive DPE 
predictions, coming from DPEMC MC simulation. 

\subsection{Exclusive $\mathbf{e^+e^-}$ and $\mathbf{\gamma \gamma}$ production}
There are SM processes in which hadrons do not dissociate in the interaction.
Without hadron dissociation, there are no underlying events then we deal with 
very clear exclusive processes.

\noindent
CDF-II studied and observed two of this exclusive channels: the exclusive production of two 
photons via QCD (gluon exchange) and the electron pair production via QED (trough two-photon exchange).

\noindent
We have observed $16$ exclusive $e^+e^-$ events with a background estimate of $2.1^{+ 0.7}_{-0.3}$. 
Each event has an $e^+e^-$ pair ($E_T(e)>$ $5$ GeV, $|\eta(e)|<$ $2$) and nothing else observable in the CDF-II detector. 
The measured cross section is $\sigma=$ $1.6^{+ 0.5}_{-0.3}$\,(stat)$\,\pm\, 0.3\,$(sys) pb, 
while the predicted cross section is $1.711 \,\pm\, 0.008$ pb. 
The kinematical properties of the events are consistent with the predictions of the LPAIR Monte Carlo. 

\noindent
We also have evidence for 3 exclusive $\gamma \gamma$ events, with a background estimate of 
$0.0^{+ 0.2}_{-0.0}$. 
Each event has two photons ($E^\gamma_T>$ $5$ GeV, $|\eta_\gamma)|<$ $1$) and nothing else observable 
in the CDF-II detector. The measured cross section for these events is  
$\sigma =$ $0.14 \,\pm \,0.14\,$(stat)$\,\pm \,0.03\,$(sys) pb and agrees with the theoretical 
prediction of $0.04$ pb with a factor $3 \div 5$ of theoretical uncertainty.

\begin{figure}[t!]
\begin{center}
\includegraphics[width=5.4cm,height=5.9cm]{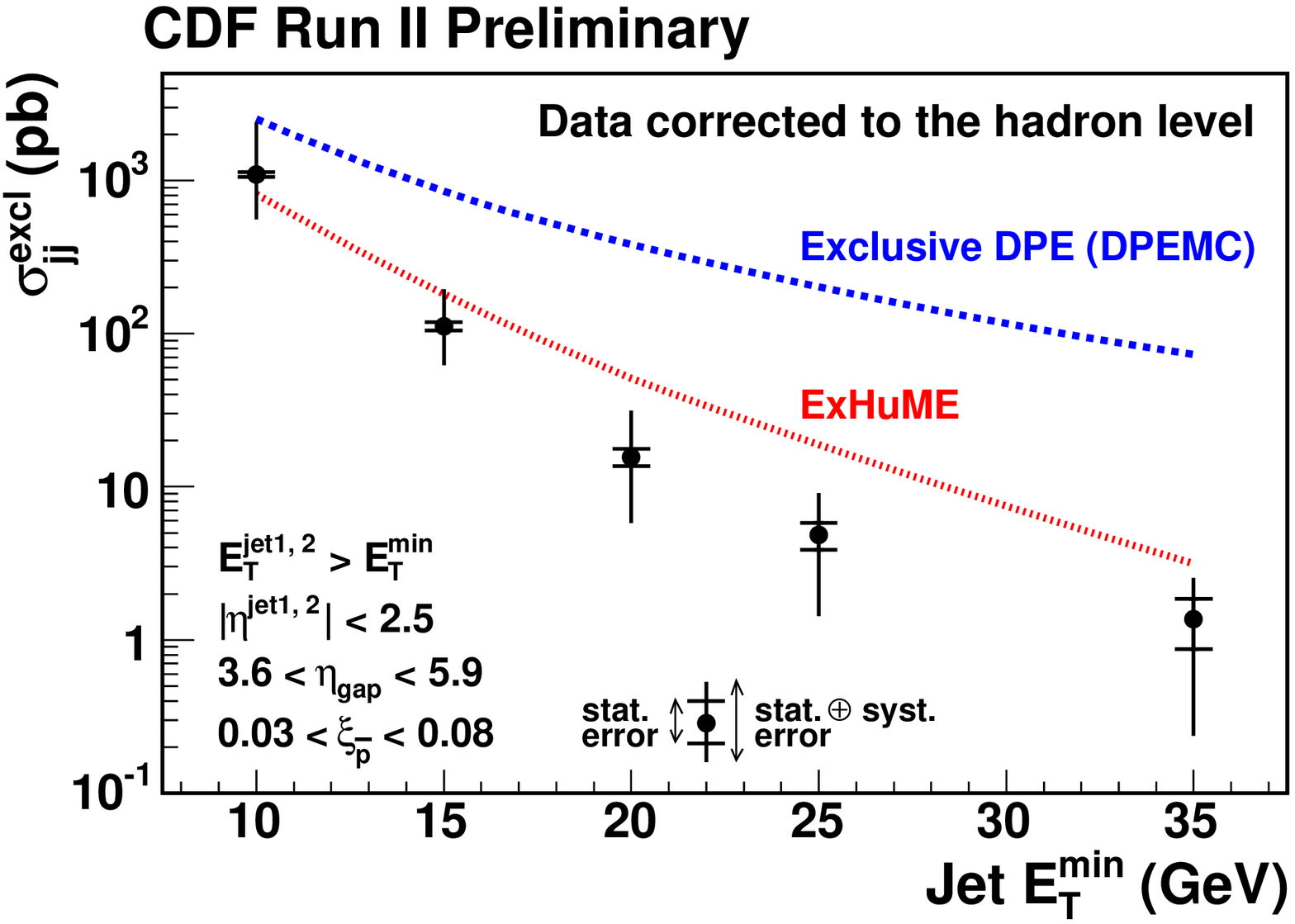}
\includegraphics[width=5.4cm,height=5.9cm]{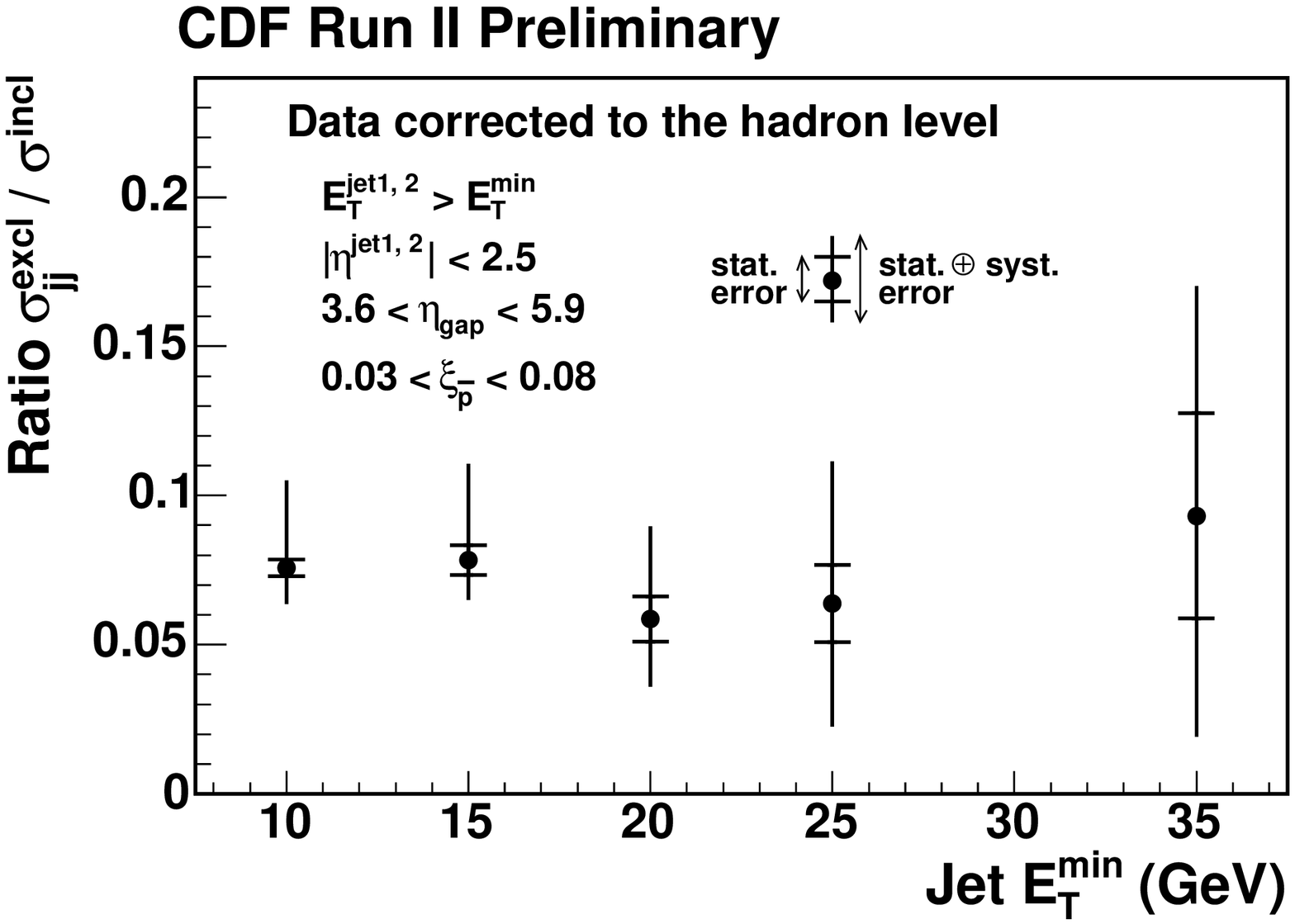}
\includegraphics[width=5.4cm,height=5.9cm]{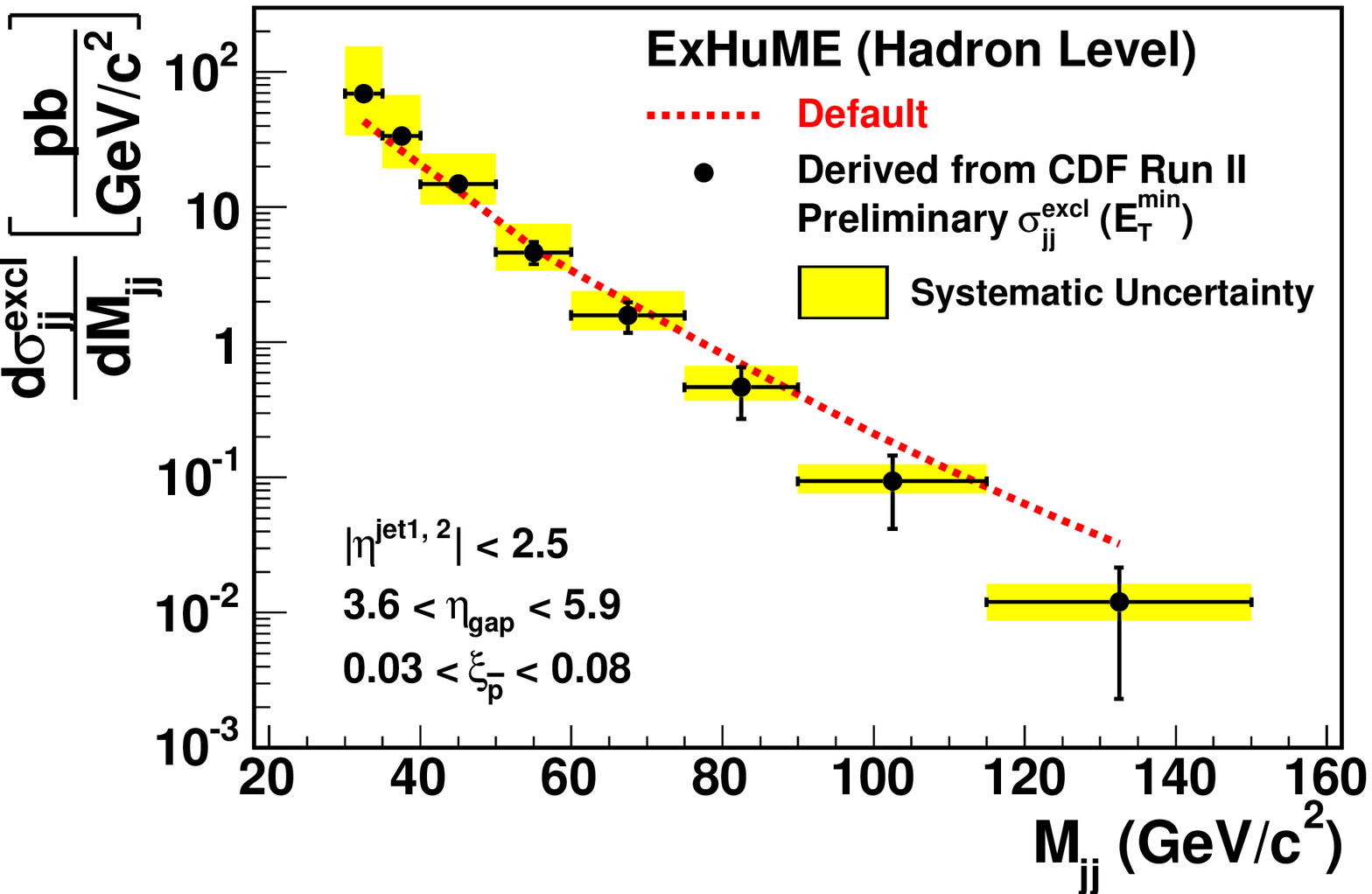}
\vspace{-1.6pc} 
\caption{\footnotesize Dijet production in DPE: 
a) Comparison with hadron level predictions of 
ExHuME (red) and Exclusive DPE in DPEMC (blue); 
b) Exclusive to inclusive dijet cross section ratio vs $E_T(min)$; 
c) ExHuME Hadron-Level differential exclusive dijet cross section vs dijet mass.}
\label{DPE}
\end{center}
\vspace{-1.7pc}
\end{figure}

\section{Summary}
CDF-II has a broad QCD analysis program: 
jets, photons, bosons + jets, heavy-flavor jets, diffractive physics.
The inclusive jet production have been measured with both $k_T$ and MidPoint
algorithm. Both the measurements are based on  ~$1$ fb$^{-1}$ of data, 
considering $5$ different rapidity regions, up to $|y|=$ $2.1$. A careful treatment 
of non perturbative effects have been taken into account. Underlying event effects
have been proved to be well under control. For central jets, $p^{jet}_T$ reach have been extended 
by ~$150$ GeV/c compared to Run I. Good agreement have been found with NLO QCD calculations.
Forward jet information can be used in future PDF global fits in order to better constrain 
the gluon PDF at high$-x$. Good agreement, with the Theoretical expectations, have been found
in all the CDF-II analysis described in this paper. We also discussed exiting observation 
such as: $e^+e^-$ production via QED exchange, $\gamma \gamma$ production via QCD exchange
and finally exclusive dijet production in DPE.

\section{Acknowledgments}
We want to thank Laszlo Jenkovszky and the Organizing Committee for their warm, 
kind and nice hospitality.

  

}

\begin{thebibliography}{99}
\bibitem{cdf_cord}
  In the CDF and  D$\not$O coordinate system, $\phi$ is the azimuthal angle and
  $\theta $ is the polar angle with respect
  to the proton beam direction. The pseudo-rapidity $\eta $ is defined as
  $\eta=- \ln \tan (\theta / 2)$. The transverse momentum of a particle is
  $p_{\rm T} = p \sin \theta $. If the magnitude of this vector is obtained
  using the calorimeter energy rather than the spectrometer momentum, it
  becomes the transverse energy $E_{\rm T}$.
  Jets are defined as clusters of energy in $\eta-\phi $ space with a
  fix cone size.
  The missing transverse energy ($\not\!\!\!E_{\rm T}$) is defined
  as the difference between the vector sum of all the transverse
  energies and zero.
\end{thebibliography}
\end{document}